\documentclass[10pt,letterpaper]{IEEEtran}

\ifCLASSINFOpdf
\else
   \usepackage[dvips]{graphicx}
\fi
\usepackage{url}

\usepackage{amsmath,graphicx,amsfonts,colortbl,subfigure,amssymb}

\definecolor{mygray}{gray}{.93}

\begin{document}

\title{SAGRNN: Self-Attentive Gated RNN for Binaural Speaker Separation with Interaural Cue Preservation}

\author{Ke Tan, Buye Xu, Anurag Kumar, Eliya Nachmani, and Yossi Adi
\thanks{K. Tan is with the Department of Computer Science and Engineering, The Ohio State University, Columbus, OH, USA (e-mail: tan.650@osu.edu). This work was done when K. Tan was a research intern at Facebook Reality Labs.}
\thanks{B. Xu and A. Kumar are with Facebook Reality Labs, Redmond, WA, USA (e-mails: \{xub, anuragkr90\}@fb.com).}
\thanks{E. Nachmani and Y. Adi are with Facebook AI Research, Israel (e-mails: \{eliyan, adiyoss\}@fb.com). E. Nachmani is also with Tel-Aviv University, Israel.}
\vspace{-1em}
}
\markboth{}
{Shell \MakeLowercase{\textit{et al.}}: Bare Demo of IEEEtran.cls for IEEE Journals}
\maketitle

\begin{abstract}
Most existing deep learning based binaural speaker separation systems focus on producing a monaural estimate for each of the target speakers, and thus do not preserve the interaural cues, which are crucial for human listeners to perform sound localization and lateralization. In this study, we address talker-independent binaural speaker separation with interaural cues preserved in the estimated binaural signals. Specifically, we extend a newly-developed gated recurrent neural network for monaural separation by additionally incorporating self-attention mechanisms and dense connectivity. We develop an end-to-end multiple-input multiple-output system, which directly maps from the binaural waveform of the mixture to those of the speech signals. The experimental results show that our proposed approach achieves significantly better separation performance than a recent binaural separation approach. In addition, our approach effectively preserves the interaural cues, which improves the accuracy of sound localization.
\end{abstract}


\IEEEpeerreviewmaketitle

\section{Introduction}
\label{sec:intro}

In real acoustic environments, a speech source of interest is frequently corrupted by interfering sounds. Human auditory system excels at attending to a target speech source, and the cocktail party problem~\cite{cherry1953some} aims to develop such capabilities in man-made devices and systems. A critical aspect of the cocktail party problem is speaker separation which aims to separate multiple concurrent speech signals of interest from a sound mixture. 

Conventionally, most of the speaker separation methods work in time-frequency (T-F) domain where T-F representations are typically computed using short-time Fourier transform (STFT). In recent years, the performance of T-F domain speaker separation has substantially improved due to the use of deep learning~\cite{hershey2016deep, isik2016single, yu2017permutation, kolbaek2017multitalker, chen2017deep, wang2018alternative, wang2018end, wang2019deep, liu2019divide}. Moreover, the advent of deep learning based speech separation has also ignited interest in time-domain approaches, which directly estimate the waveform of clean speech from that of the mixture without resorting to a T-F representation. A notable time-domain speaker separation approach is TasNet~\cite{luo2019conv}, which yields comparable scale-invariant signal-to-noise ratios (SI-SNRs) and signal-to-distortion ratios (SDRs) to the ideal ratio mask (IRM). Other related studies include~\cite{stoller2018wave, venkataramani2018end, luo2020dual, zhang2020furcanext, nachmani2020voice} and~\cite{zeghidour2020wavesplit}.


\begin{figure*}[t]
	\vspace{-20pt}
	\centering
	\includegraphics[width=14cm]{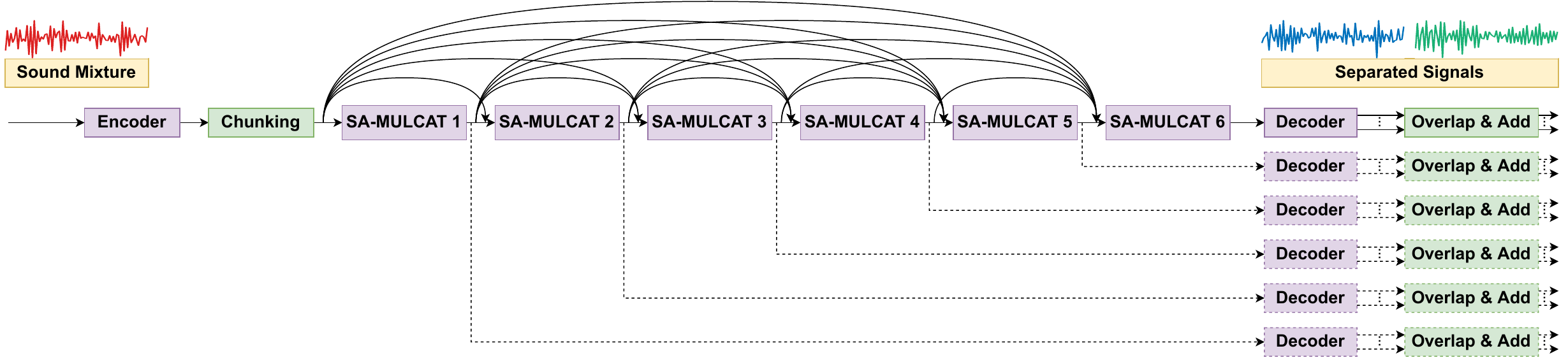}
	\vspace{-8pt}
	\caption{(Color Online). Diagram of SISO SAGRNN. The dotted lines represent the procedures that exist only during training.}
	\label{fig:siso_sagrnn}
	\vspace{-10pt}
\end{figure*}

While several time-domain monaural speaker separation methods have been developed, very few works have focused on binaural separation. Moreover, most existing binaural separation systems have a multiple-input single-output (MISO) layout, which produce a mono estimate for each of the target speakers from a binaural mixture~\cite{zohourian2016binaural, liu2018iterative, dadvar2019robust}. Hence these systems do not preserve interaural cues such as interaural time differences (ITDs) and interaural level differences (ILDs), which are crucial for human listeners to perform sound localization and lateralization~\cite{domnitz1977lateral, hershkowitz1969interaural}. 

On the T-F domain front, various techniques have been developed to preserve binaural cues in the estimated signals. One can apply a common real-valued T-F mask or spectral gain to both left and right channels~\cite{mandel2009model, zohourian2018gsc}. Alternatively, binaural cues can be preserved by applying adaptive beamformers with additional constraints that encourage interaural cue preservation~\cite{van2007binaural, marquardt2015theoretical, hadad2015theoretical}. However, these techniques sacrifice the separation performance. More recently, a multiple-input multiple-output (MIMO) TasNet~\cite{han2020real} was designed, which produces a binaural estimate for each speaker. MIMO TasNet yields significantly better speech quality than the single-channel TasNet while preserving both ITDs and ILDs. 

In this paper, we propose a novel framework called \emph{multiple-input multiple-output self-attentive gated recurrent neural network} (MIMO SAGRNN) for binaural speaker separation. The proposed SAGRNN network architecture extends the gated RNN in~\cite{nachmani2020voice} by additionally incorporating self-attention mechanisms and dense connectivity (DC). We then derive MIMO SAGRNN from a single-input single-output (SISO) SAGRNN by first extending the SISO SAGRNN into a multiple-input single-output (MISO) layout by creating two encoders, one for the \emph{reference ear} input and the other for the \emph{non-reference ear} input. This MISO SAGRNN estimates the separated signals in the reference ear. The MIMO system is formulated by alternately treating each ear as the reference ear, yielding estimates for both ears in a symmetric manner.

The rest of this paper is organized as follows. Section~\ref{sec:algorithm} describes our proposed approach. The experimental results are presented in Section~\ref{sec:exp}, and Section~\ref{sec:conclusion} concludes this paper.

\vspace{-0.2cm}
\section{Algorithm Description}
\label{sec:algorithm}



We progressively develop a MIMO system for binaural speaker separation. Specifically, we start with a SISO SAGRNN architecture, and then present the MIMO setup. 


\vspace{-10pt}
\subsection{SISO SAGRNN}
As in~\cite{nachmani2020voice}, the separation framework of a SISO SAGRNN comprises three stages: \emph{encoding and chunking}, \emph{block processing}, and \emph{decoding and overlap-add}. A time-domain input mixture is transformed into a set of overlapped chunks via encoding and chunking, which leads to a 3-D embedding. Subsequently, the 3-D embedding is passed into stacked RNN blocks to perform intra-chunk (local) and inter-chunk (global) modeling alternately and iteratively. The 3-D representation learned by the last RNN block is decoded and then transformed back to the time domain by an overlap-add operator.

Given a $T$-sample input waveform $y \in \mathbb{R}^T$, an encoder is used to segment and encode $y$ into $L$ overlapped time frames with a frame size of $P$ and a hop size of $P/2$, yielding a 2-D embedding $\mathbf{U} \in \mathbb{R}^{N \times L}$. Specifically, the encoder consists of a 1-D strided convolutional layer with $N$ output channels, followed by a rectified linear activation function. We divide the time frames into $S$ overlapped chunks with a chunk size of $R$ and a hop size of $R/2$. These chunks are then concatenated into a 3-D embedding $\widetilde{\mathbf{W}} = [\mathbf{W}_1, \dots, \mathbf{W}_S] \in \mathbb{R}^{N \times S \times R}$, where $\mathbf{W}_1, \dots, \mathbf{W}_S \in \mathbb{R}^{N \times R}$ are the 2-D chunks.

\begin{figure*}[t]
	\vspace{-20pt}
	\centering
	\includegraphics[width=16cm]{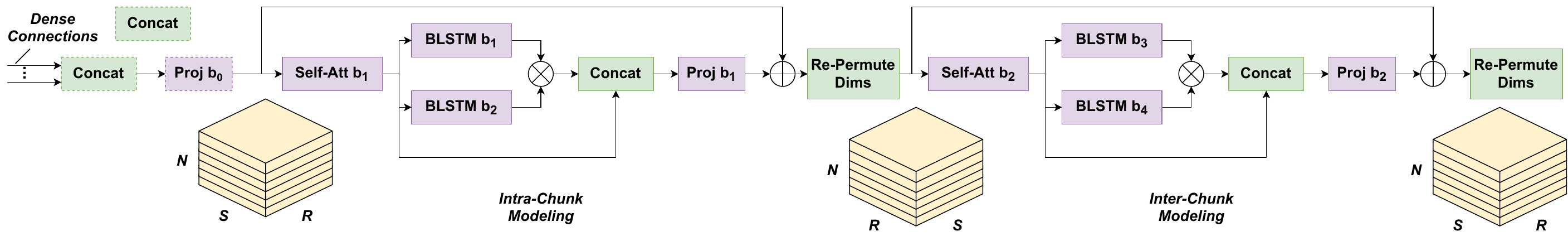}
	\vspace{-8pt}
	\caption{(Color Online). Diagram of the SA-MULCAT block. The dotted lines indicate that the procedures exist in all SA-MULCAT blocks except the first one. The symbol $\bigotimes$ represents the element-wise multiplication, and $\bigoplus$ the element-wise addition.}
	\label{fig:SA_MULCAT}
	\vspace{-15pt}
\end{figure*}

\begin{figure}[t]
	\vspace{-5pt}
	\centering
	\subfigure[]{%
		\centering
		\includegraphics[width=3.0cm]{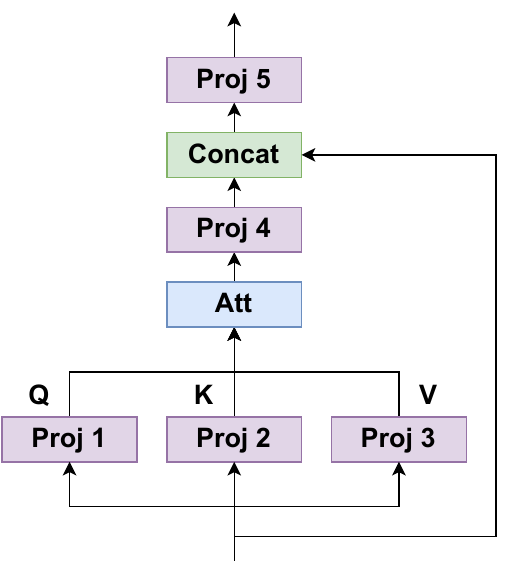}
		\label{fig:subfig:self-attention}
	}\hspace{1mm}%
	\subfigure[]{%
		\centering
		\includegraphics[width=5.5cm]{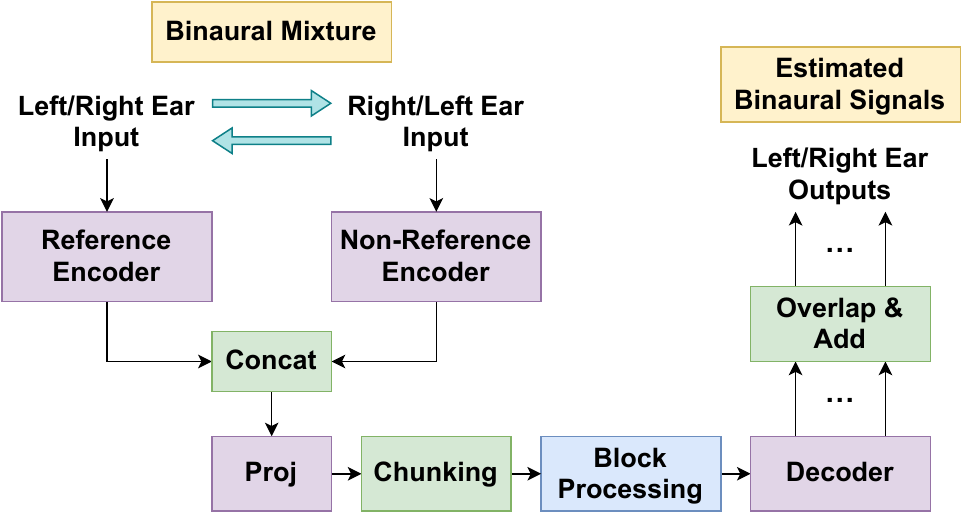}
		\label{fig:subfig:miso_mimo}
	}%
	\vspace{-5pt}
	\caption{(Color Online). (a) Diagram of the self-attention block. (b) Overview of the MIMO separation system with an underlying MISO system.}
	\label{fig:sa_mimo}
	\vspace{-10pt}
\end{figure}

Subsequently, the 3-D embedding $\widetilde{\mathbf{W}}$ is fed into a series of $B$ RNN blocks for processing. To improve the information and gradient flow between blocks, we propose a dense connectivity pattern: each block receives the outputs of all preceding blocks, i.e. $\widetilde{\mathbf{W}}_b = \mathcal{H}_b ([\widetilde{\mathbf{W}}_0, \dots, \widetilde{\mathbf{W}}_{b-1}])$ for $b = 1, \dots, B$,
where $\mathcal{H}_b$ denotes the mapping function defined by the $b$-th block, and $\left[\cdot, \dots, \cdot \right]$ the concatenation operation. The output embedding of the $b$-th block is represented by $\widetilde{\mathbf{W}}_b$, where $\widetilde{\mathbf{W}}_0 = \widetilde{\mathbf{W}}$ and $\widetilde{\mathbf{W}}_b \in \mathbb{R}^{N \times S \times R}, \forall b$. The dense connections encourage feature reuse among blocks, which explicitly leverage different information learned by different blocks.

Similar to~\cite{nachmani2020voice}, we use a multi-scale loss for training, which necessitates producing a waveform estimate for each speaker 
after each block. We decode the output embedding of each block with a decoder, which comprises a parametric rectified linear function~\cite{he2015delving} followed by a 2-D 1$\times$1 convolutional layer with $C\cdot N$ output channels. The decoded feature of size $CN \times S \times R$ is divided into $C$ 3-D representations of size $N \times S \times R$, corresponding to the $C$ speech sources. These 3-D representations are transformed back to waveforms by two successive overlap-add operations at the chunk level and the frame level, respectively. Note that the same decoder is applied to the output of each block. Fig.~\ref{fig:siso_sagrnn} depicts the SISO SAGRNN. 

A series of multiply-and-concatenate (MULCAT) blocks are employed to model the intra-chunk and inter-chunk dependencies. In this study, we extend the MULCAT block by introducing self-attention~\cite{vaswani2017attention}, which amounts to a self-attention based MULCAT (SA-MULCAT) block illustrated in Fig.~\ref{fig:SA_MULCAT}. The concatenation of paths from the dense connections is fed into a linear projection layer for dimension reduction, yielding an embedding of size $N \times S \times R$. The resulting embedding is successively passed through two subblocks, one for intra-chunk modeling and the other for inter-chunk modeling. In each subblock, we employ a self-attention block followed by a gated RNN module, which consists of two bidirectional long short-term memory (BLSTM) layers coupled with each other. Each BLSTM contains $H$ units in each direction. The Hadamard product of their outputs is concatenated with the input to the gated RNN module, and then passed into a linear projection layer for dimension reduction. In addition, a skip connection is used to bypass the subblock. After the first subblock, the dimensions of the 3-D representation are re-permuted, so that sequential modeling can be performed across chunks in the second subblock. After the second subblock, the dimensions are re-permuted back.

The self-attention block is illustrated in Fig.~\ref{fig:subfig:self-attention}. We first divide a 3-D representation into a set of 2-D slices $\mathbf{Z} \in \mathbb{R}^{M \times N}$, where $M = R$ for intra-chunk modeling and $M = S$ for inter-chunk modeling. Each slice is linearly projected to a query matrix $\mathbf{Q}$, a key matrix $\mathbf{K}$ and a value matrix $\mathbf{V}$ by three different projection layers, where $\mathbf{Q}, \mathbf{K}, \mathbf{V} \in \mathbb{R}^{M \times D}$ and $D$ is set to 64. We apply a scaled dot-product attention function:
\vspace{-6pt}
\begin{equation}
Attention(\mathbf{Q}, \mathbf{K}, \mathbf{V}) = SoftMax(\frac{\mathbf{Q}\mathbf{K}^\top}{\sqrt{D}}) \mathbf{V},
\vspace{-6pt}
\end{equation}
where $SoftMax(\cdot)$ denotes the softmax function across columns. The output of the attention function is computed as a weighted sum of the values, where the weight assigned to each value is derived by measuring the similarities between the queries and the keys. Subsequently, all the attention output slices are merged and then linearly projected back to the size of the input 3-D representation. With a skip connection, this representation is concatenated with the input to the self-attention block, and then projected back to the original size. The use of self-attention is motivated by its recent success on monaural speech enhancement and dereverberation~\cite{koizumi2020speech, zhao2020monaural}, which has demonstrated its capability of capturing long-term dependencies in target speech and interference. By leveraging the relevance among features at different time steps, self-attention produces a dynamic representation in adapting to different acoustic conditions.

\vspace{-6pt}
\subsection{MIMO SAGRNN}
As shown in Fig.~\ref{fig:subfig:miso_mimo}, a \emph{reference encoder} and a \emph{non-reference encoder} are employed to process the binaural mixture waveforms. The resulting 2-D embeddings are concatenated and then linearly projected to the size of $N \times L$. Subsequently, we successively perform block processing, decoding and overlap-add, akin to the SISO system. In this MISO system, the separation outputs always correspond to the reference ear. We formulate the MIMO system by alternately treating each ear as the reference. Specifically, the separation outputs for the left ear are obtained by treating the left ear as the reference ear and the right ear as the non-reference. The separation outputs for the right ear are obtained by swapping the inputs of the two ears. Note that the same MISO system is used for separation in both channels. Such a cross-ear referencing strategy selects the target channel by exploiting discriminative information within the ordered pair of channels.

\vspace{-10pt}
\subsection{Training Objective}
We use the plain SNR rather than the widely-used SI-SNR~\cite{luo2019conv} as the training objective. The rationale is that SI-SNR training cannot preserve the ILD in the binaural estimates, as the power scale of the estimated signals is insusceptible to training due to the scale invariance. 
The mean of the SNR losses from all SA-MULCAT blocks is used for training. The waveforms of the clean speech signals are used as the training target for calculating the losses from all blocks. In addition, we apply the permutation invariant training~\cite{kolbaek2017multitalker} criterion to the loss from each block individually, which allows the label permutation to change from one block to another.

\vspace{-0.2cm}
\section{Experiments}
\label{sec:exp}

\subsection{Experimental Setup}
We simulate a noise-free dataset and a noisy dataset from the WSJ0-2mix dataset~\cite{hershey2016deep}, which contains 20,000, 5,000 and 3,000 mixtures in the training, validation and testing sets, respectively. For both datasets, we convolve each pair of utterances in WSJ0-2mix with two randomly sampled head-related impulse responses (HRIRs) from the CIPIC HRTF Database~\cite{algazi2001cipic} respectively, which contains 45 subjects with 25~(azimuths)~$\times$~50~(elevations) directions for each subject. Specifically, we choose 35 subjects for training and cross validation, and use the 10 remaining subjects for testing. For the noisy dataset, we additionally simulate uncorrelated noise sources by randomly selecting HRIRs for them, where the number of noise sources is randomly sampled between 1 and 10. Note that all sound sources are placed in different directions. We use a set of roughly 65,000 noises from the DNS Challenge~\cite{reddy2020interspeech} for training and cross validation, and a different set~\cite{Robinhood76} of roughly 1,300 noises for testing. The SNR (w.r.t. the speech mixture in the left ear) is randomly chosen between -10~dB and 10~dB. All signals are sampled at 8~kHz.


We train the models on 4-second segments with the AMSGrad optimizer~\cite{reddi2018convergence} with a minibatch size of 4. The learning rate is initialized to 0.0002, which decays by 0.98 every 2 epochs. Gradient clipping with a maximum $\ell^2$ norm of 3 is applied during training. The network hyperparameters for MIMO SAGRNN are as follow: $P=8$, $N=128$, $R=126$, $H=128$, $D=64$ and $B=6$. Note that the value of $R$ is selected such that $R \approx S = 128$ for the training segments.

We use several monaural and binaural separation models for comparison; the monaural models are trained and evaluated on each ear individually. Specifically, we use TasNet~\cite{luo2019conv}, dual-path RNN (DPRNN)~\cite{luo2020dual} and the gated RNN in~\cite{nachmani2020voice} as monaural baselines. We use MIMO TasNet~\cite{han2020real} as a binaural baseline. We slightly adjust the hyperparameter configurations of all baselines, so that they have comparable model sizes to our MIMO SAGRNN. For the noncausal temporal convolutional network (TCN) in TasNet~\cite{luo2019conv}, the number of repeated stacks is set to 4. For DPRNN, the number of output channels in the encoder and the decoder is set to 128, and the number of units in each direction for each BLSTM to 200. For MIMO TasNet, we replace the causal TCN by a noncausal TCN, with bottleneck size of 128. In addition, the number of output channels in the encoder and the decoder of MIMO TasNet is set to 512.

\vspace{-8pt}
\subsection{Experimental Results}
\subsubsection{Separation Results and Analysis}
\begin{table}[t]
	\vspace{-8pt}
	\caption{Comparison of different systems in the noise-free condition.}
	\vspace{-6pt}
	\centering
	\resizebox{0.48\textwidth}{!}{
		\begin{tabular}{|l|c|c|c|c|c|}
			\hline
			Metrics & \# param. & $\Delta$SDR (dB) & $\Delta$SNR (dB) & ESTOI (\%) & PESQ \\
			\hline
			Mixture & - & 0.00 & 0.00 & 56.10 & 1.99 \\
			\hline
			TasNet & 6.66~M & 15.78 & 15.80 & 90.25 & 3.30 \\ 
			DPRNN & 6.99~M & 18.25 & 18.28 & 92.64 & 3.49 \\
			Gated RNN & 7.56~M & 18.37 & 18.41 & 92.52 & 3.52 \\
			Oracle IBM & - & 13.69 & 13.61 & 88.49 & 3.36 \\
			Oracle IRM & - & 13.05 & 13.07 & 93.33 & 3.73 \\
			Oracle PSM & - & 16.77 & 16.58 & 95.55 & 3.91 \\
			\hline
			MIMO TasNet & 7.32~M & 21.14 & 20.69 & 95.53 & 3.73 \\
			Oracle MB-MVDR & - & 17.13 & 10.44 & 95.77 & 3.66 \\
			\hline
			\rowcolor{mygray}
			MIMO SAGRNN & 8.71~M & \textbf{27.19} & \textbf{26.88} & \textbf{98.08} & \textbf{4.06} \\		 
			\hspace{1mm}$-$ half multi-scale loss~(i) & 8.71~M & 24.35 & 23.99 & 96.70 & 3.90 \\
			\hspace{1mm}$-$ multi-scale loss~(ii) & 8.71~M & 22.93 & 22.45 & 95.69 & 3.79 \\
			\hspace{1mm}$-$ DC~(iii) & 8.38~M & 24.02 & 23.64 & 96.54 & 3.87 \\
			\hspace{1mm}$-$ SA~(iv) & 7.92~M & 23.13 & 22.75 & 96.12 & 3.82 \\
			\hspace{1mm}$-$ DC $-$ SA~(v) & 7.59~M & 21.97 & 21.64 & 95.66 & 3.73 \\
			\hline
		\end{tabular}
	}
	\label{tab:noise_free}
	\vspace{-11pt}
\end{table}

Tables~\ref{tab:noise_free} and~\ref{tab:noisy} show comparison among different approaches for the noise-free and noisy conditions respectively. The separation results are reported in terms of SDR improvement ($\Delta$SDR), SNR improvement ($\Delta$SNR), extended short-time objective intelligibility (ESTOI)~\cite{jensen2016algorithm}, and perceptual evaluation of speech quality (PESQ)~\cite{rix2001perceptual}. We can observe that MIMO TasNet produces consistently better results than the monaural baselines. Moreover, our proposed MIMO SAGRNN substantially outperforms MIMO TasNet in all the four metrics. For the noise-free condition, MIMO SAGRNN improves SDR by 6.05~dB and SNR by 6.19~dB over MIMO TasNet. Some demos can be found at https://jupiterethan.github.io/sagrnn.github.io/.

\begin{table}[t]
	\vspace{-24pt}
	\caption{Comparison of different systems in the noisy condition.}
	\vspace{-5pt}
	\centering
	\resizebox{0.4\textwidth}{!}{
		\begin{tabular}{|l|c|c|c|c|c|}
			\hline
			Metrics & $\Delta$SDR (dB) & $\Delta$SNR (dB) & ESTOI (\%) & PESQ \\
			\hline
			Mixture & 0.00 & 0.00 & 27.80 & 1.49 \\
			\hline
			TasNet & 11.71 & 13.33 & 54.65 & 2.12 \\ 
			DPRNN & 11.91 & 13.55 & 53.72 & 2.17 \\
			Gated RNN & 12.89 & 14.27 & 59.09 & 2.30 \\
			Oracle IBM & 14.11 & 14.64 & 71.19 & 2.63 \\
			Oracle IRM & 13.19 & 13.97 & 84.22 & 3.33 \\
			Oracle PSM & 17.41 & 17.62 & 89.40 & 3.55 \\
			\hline
			MIMO TasNet & 14.40 & 15.23 & 63.79 & 2.41 \\
			Oracle MB-MVDR & 4.98 & 4.90 & 42.71 & 1.79 \\
			\rowcolor{mygray}
			MIMO SAGRNN & \textbf{17.53} & \textbf{17.95} & \textbf{75.14} & \textbf{2.78} \\		 
			\hline
		\end{tabular}
	}
	\label{tab:noisy}
	\vspace{-5pt}
\end{table}

In addition, we compare MIMO SAGRNN with several oracle approaches, including ideal binary mask (IBM), ideal ratio mask (IRM), phase-sensitive mask (PSM)~\cite{erdogan2015phase} and an oracle masking-based minimum variance distortionless response (MB-MVDR) beamformer. We use an open-source implementation~\cite{beamformers} of the oracle MB-MVDR beamformer, with a frame length of 64~ms and a frame shift of 32~ms. The IRM is used to calculate the spatial covariance matrices. We alternately treat each channel as the reference channel to produce the binaural estimate. As shown in Table~\ref{tab:noise_free}, our approach consistently outperforms the ideal masks and the oracle beamformer in the noise-free condition. In the noisy condition (Table~\ref{tab:noisy}), our approach produces slightly higher SDR and SNR but lower ESTOI and PESQ than the PSM. Note that, ESTOI and PESQ improvements over the mixtures using the oracle MB-MVDR beamformer dramatically decrease in the noisy condition compared with the noise-free condition. This is likely because the directionality of the sound sources is smeared due to the presence of multiple noise sources. In contrast, our approach is more robust against the noise field.

\subsubsection{Ablation Study}
We conduct an ablation study to understand the contribution of each component in our approach. Several variants of MIMO SAGRNN are compared in Table~\ref{tab:noise_free}: (i) using multi-scale loss computed from only the last three RNN blocks; (ii) using loss computed from only the last RNN block; (iii) without dense connections; (iv) without self-attention blocks; (v) without dense connections and self-attention blocks. It is shown that self-attention and dense connectivity are crucial for MIMO SAGRNN. Without self-attention, for example, SDR decreases by 4.06~dB and SNR by 4.13~dB. We also compare the SISO and MIMO SAGRNNs on the better-ear channel; {\it better ear} is defined as the ear that is closer to the target speech source. The azimuth position of the speech source is used to determine the better ear. Table~\ref{tab:better_ear} shows that MIMO SAGRNN yields significantly better results than SISO SAGRNN and the gated RNN in~\cite{nachmani2020voice} on all metrics, evidencing that the binaural inputs are effectively leveraged by our MIMO system.

\begin{table}[t]
	\vspace{-5pt}
	\caption{Comparison between SISO and MIMO SAGRNNs on the better-ear channel for the noisy condition.}
	\vspace{-5pt}
	\centering
	\resizebox{0.4\textwidth}{!}{
		\begin{tabular}{|l|c|c|c|c|c|}
			\hline
			Metrics & $\Delta$SDR (dB) & $\Delta$SNR (dB) & ESTOI (\%) & PESQ \\
			\hline
			Mixture & 0.00 & 0.00 & 31.84 & 1.57 \\
			\hline
			Gated RNN & 12.59 & 13.50 & 64.64 & 2.42\\
			SISO SAGRNN & 13.19 & 14.06 & 66.91 & 2.49 \\ 
			MIMO SAGRNN & 15.94 & 16.32 & 75.66 & 2.79 \\		 
			\hline
		\end{tabular}
	}
	\label{tab:better_ear}
	\vspace{-14pt}
\end{table}

\begin{table}[t]
	\vspace{-24pt}
	\caption{Evaluation of interaural cue preservation with binaural sound localization for the noise-free condition.}
	\vspace{-5pt}
	\centering
	\resizebox{0.48\textwidth}{!}{
		\begin{tabular}{|l|c|c|c|c|c|}
			\hline
			Metrics & $\Delta$Azimuth ($^\circ$) & $\Delta$ITD ($\mu s$) & \multicolumn{3}{c|}{$\Delta$ILD (dB)} \\
			\hline
			Frequency Channels & - & - & 2.07~kHz & 3.08~kHz & 3.75~kHz \\
			\hline
			Mixture & 26.03 & 255.85 & 4.31 & 4.85 & 5.12 \\
			\hline
			TasNet & 14.40 & 25.71 & 0.62 & 0.68 & 0.95 \\	
			DPRNN & 13.64 & 25.71 & 0.62 & 0.68 & 0.93 \\
			Gated RNN & 13.82 & 23.42 & 0.91 & 0.73 & 1.22 \\
			Oracle IBM & 12.25 & 15.75 & 0.90 & 0.98 & 1.16 \\
			Oracle IRM & 9.43 & 55.00 & 0.29 & 0.28 & 0.39 \\
			Oracle PSM & 4.95 & 16.20 & 0.37 & 0.40 & 0.55 \\
			\hline
			MIMO TasNet & 6.45 & 20.10 & 0.88 & 0.77 & 1.13 \\
			Oracle MB-MVDR & 34.05 & 371.35 & 6.31 & 7.06 & 7.39 \\
			\rowcolor{mygray}
			MIMO SAGRNN & \textbf{5.88} & \textbf{14.95} & \textbf{0.53} & \textbf{0.45} & \textbf{0.70} \\	
			\hline
		\end{tabular}
	}
	\label{tab:localization}
	\vspace{-10pt}
\end{table}

\subsubsection{Evaluation of Interaural Cue Preservation}

\begin{figure}[t]
    \vspace{-5pt}
	\centering
	\subfigure[ITD Dist. (Clean)]{%
		\centering
		\includegraphics[width=4.25cm]{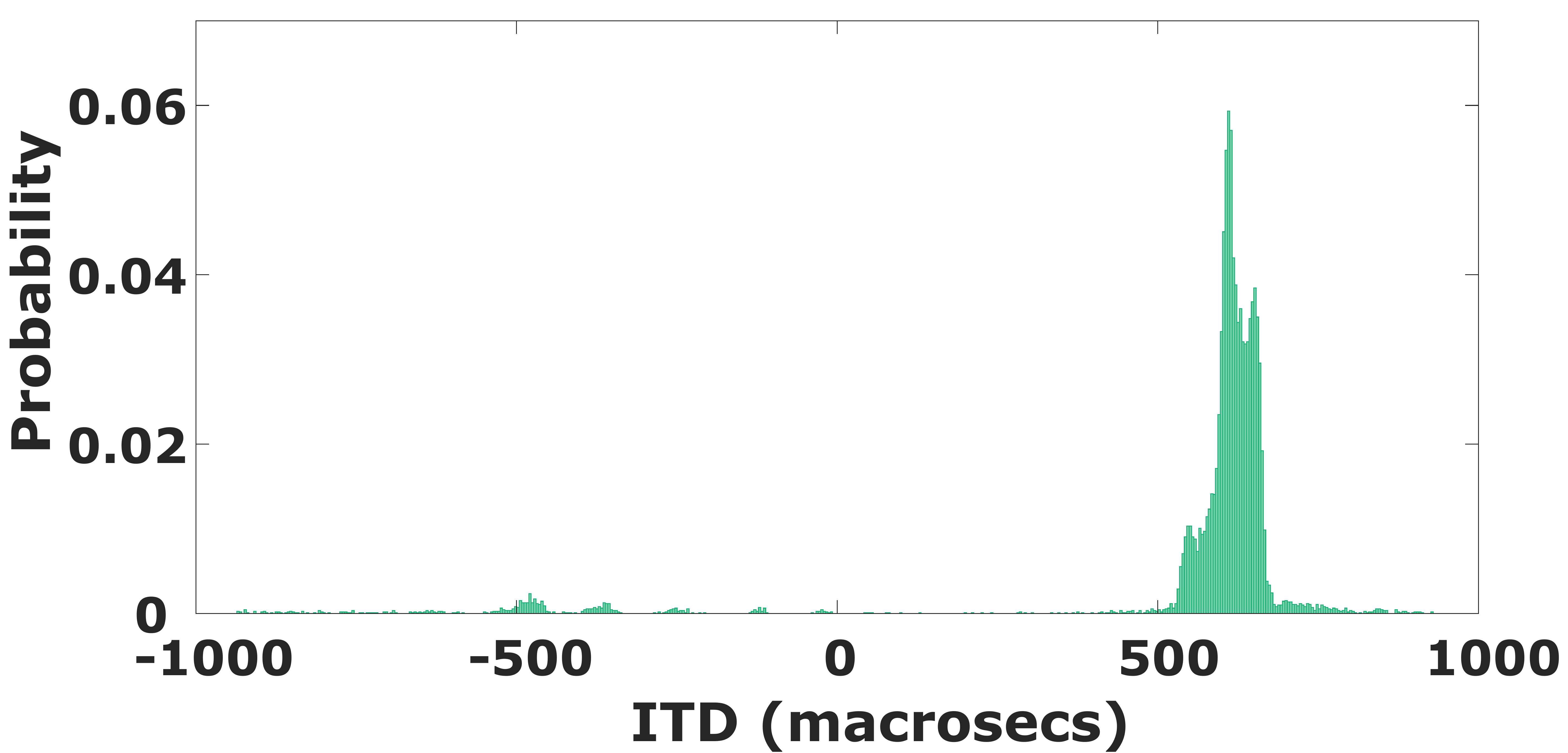}
		\label{fig:subfig:itd_sphs}
	}%
	\subfigure[ITD Dist. (Estimated)]{%
		\centering
		\includegraphics[width=4.25cm]{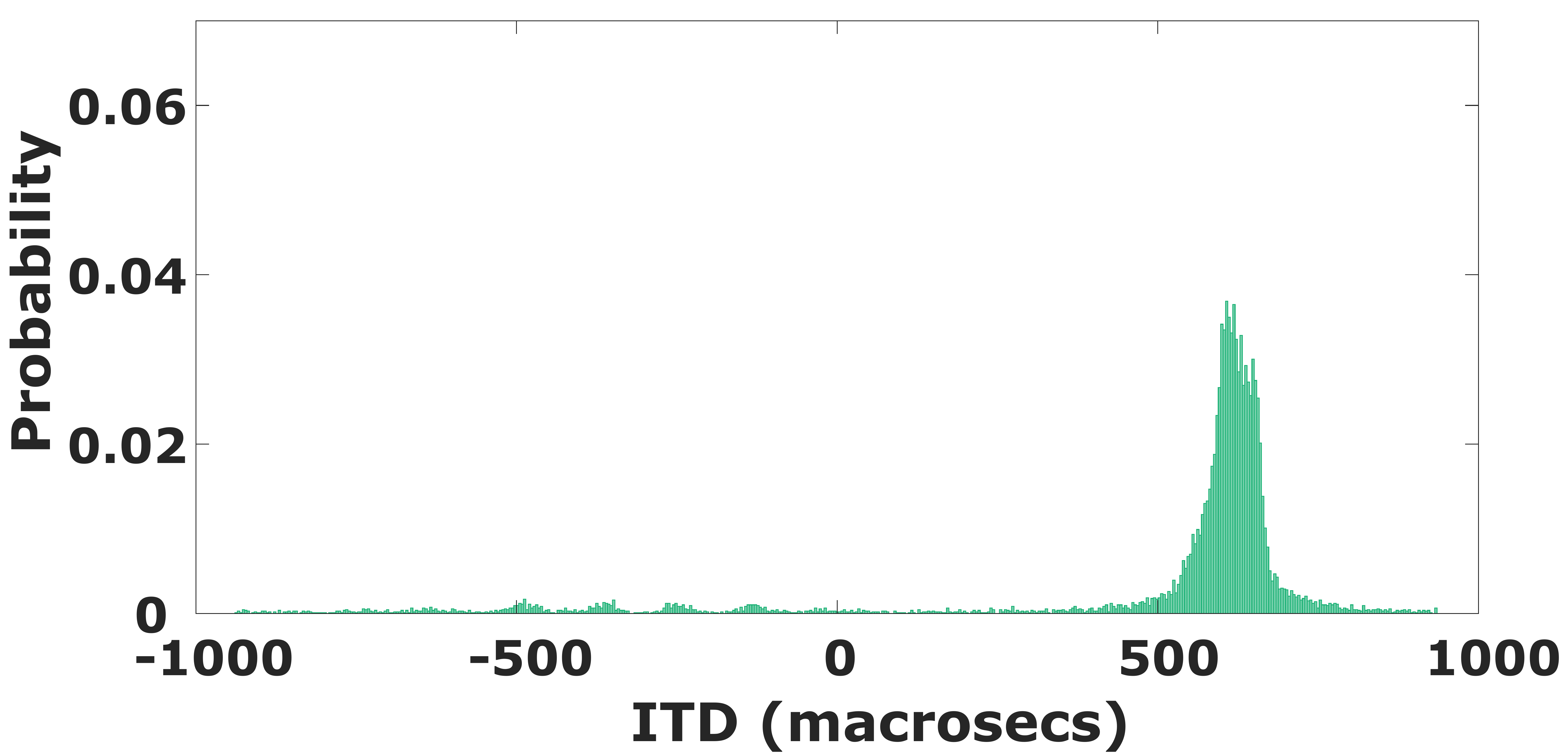}
		\label{fig:subfig:itd_sphs_est}
	}%
	
	\vspace{-8pt}
	\subfigure[ILD Dist. at 3.75~kHz (Clean)]{%
		\centering
		\includegraphics[width=4.25cm]{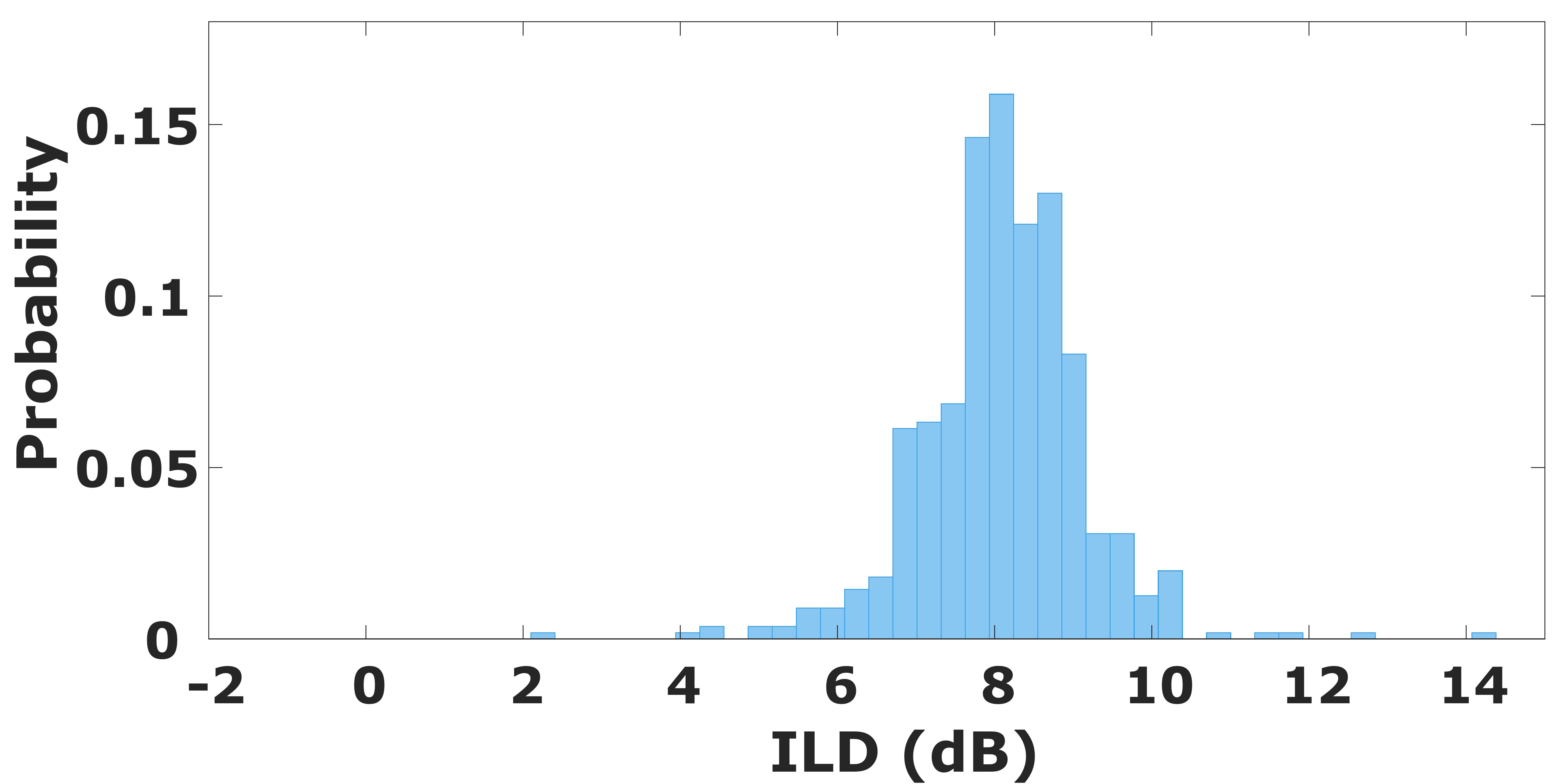}
		\label{fig:subfig:ild_sphs}
	}%
	\subfigure[ILD Dist. at 3.75~kHz (Estimated)]{%
		\centering
		\includegraphics[width=4.25cm]{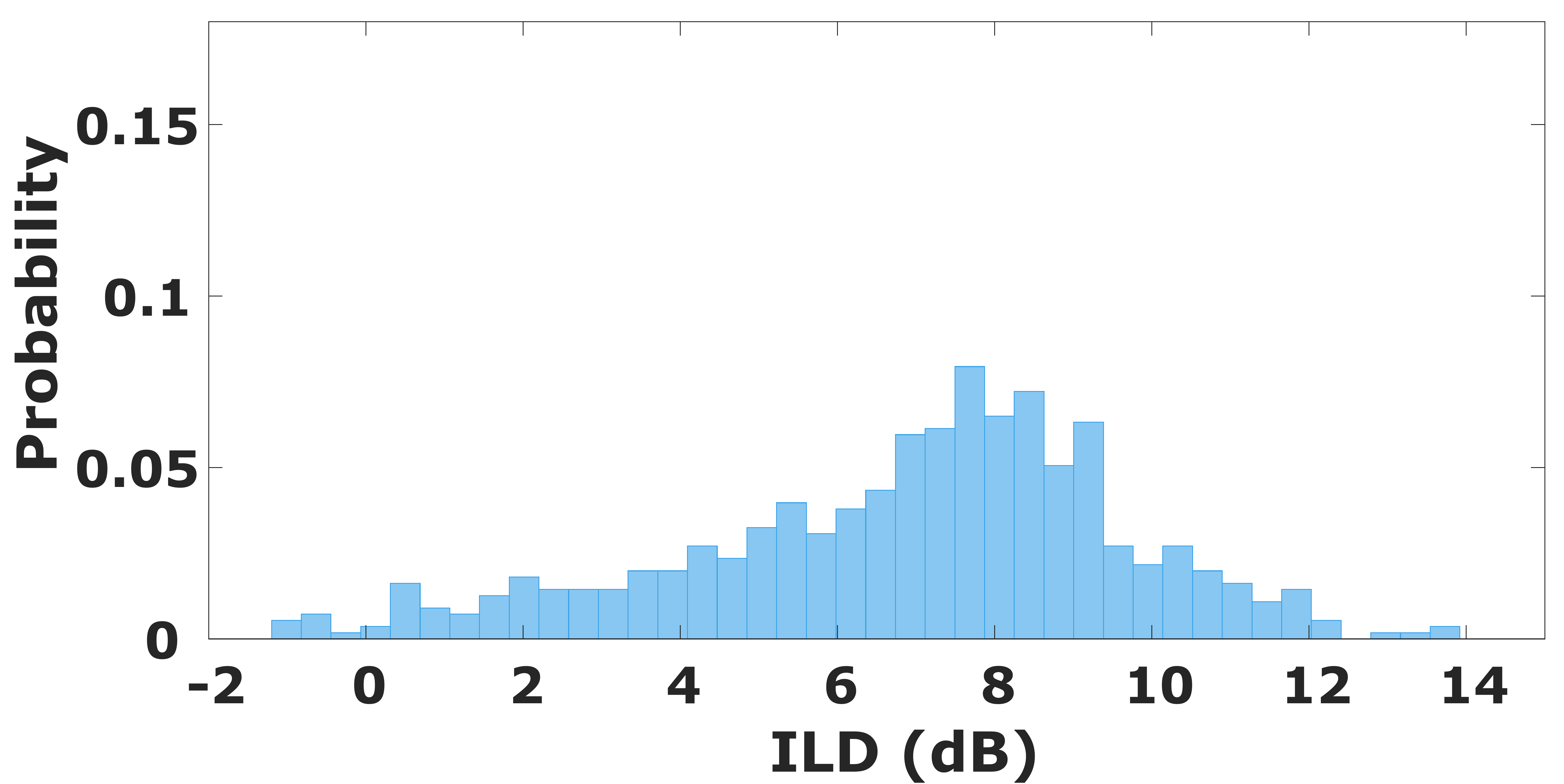}
		\label{fig:subfig:ild_sphs_est}
	}%
	
	\vspace{-5pt}
	\caption{(Color Online). An example of ITD and ILD distributions.}
	\label{fig:hist_itd_ild}
	\vspace{-15pt}
\end{figure}

Lastly, we evaluate the preservation of interaural cues in the estimated binaural signals under noise-free condition. We apply a binaural sound localization algorithm~\cite{may2011probabilistic} to the binaural estimates, of which an open-source implementation is available. This implementation estimates the azimuth position of the sound source at the frame level, as well as the ITD and the ILD for each T-F unit of a 32-channel cochleagram based on a gammatone filterbank. The average frame-level azimuth errors are presented in Table~\ref{tab:localization}. Given the dominance of the ITD cue at low frequencies (below 1.5~kHz) in sound localization~\cite{wightman1992dominant}, we only take into account the frequency bands corresponding to gammatone filters with a maximum center frequency of approximately 1.5~kHz. Since ILD is highly frequency-dependent due to diffraction and attenuation of the sounds, we calculate the average ILD errors individually for three empirically selected frequency channels, corresponding to the gammatone filters with the center frequencies of roughly 2.07, 3.08 and 3.75~kHz. Given the fact that all sound sources are stationary in this study, we summarize only one ITD/ILD from an entire utterance in the following way. We plot a histogram of the T-F unit level ITDs/ILDs, and then estimate the ITD/ILD based on the center value of the highest bin. The number of bins is empirically set to 500 for ITD and 40 for ILD. An example of ITD and ILD histograms are presented in Fig.~\ref{fig:hist_itd_ild}. As shown in Table~\ref{tab:localization}, our approach reduces the azimuth error by 20.15$^\circ$ compared to the mixtures. Moreover, our approach yields consistently smaller azimuth, ITD and ILD errors than MIMO TasNet and the monaural baselines, showing that our approach preserves the interaural cues more effectively.

\vspace{-4pt}
\section{Conclusion}
\label{sec:conclusion}
\vspace{-1pt}
We have proposed an end-to-end MIMO system for binaural speaker separation with interaural cue preservation. We developed a novel framework which relies on self-attention and dense connectivity for improved speaker separation. Our experimental results show that the proposed approach significantly outperforms a binaural separation approach (i.e. MIMO TasNet) in terms of $\Delta$SDR, $\Delta$SNR, ESTOI and PESQ. Moreover, our approach effectively preserves the auditory spatial cues of talkers. For future work, we would devote more efforts to the design of MIMO systems for real-time processing, as well as exploring binaural speaker separation in more realistic acoustic conditions (e.g. with reverberation and diffuse noise).


\newpage
\bibliographystyle{abbrv}
\bibliography{./bare_jrnlV2}

\begin{thebibliography}{10}

\bibitem{beamformers}
Beamformers.
\newblock \url{https://pypi.org/project/beamformers}.

\bibitem{Robinhood76}
Robinhood76 sounds.
\newblock \url{https://freesound.org/people/Robinhood76/}.

\bibitem{algazi2001cipic}
V.~R. Algazi, R.~O. Duda, D.~M. Thompson, and C.~Avendano.
\newblock The {CIPIC} {HRTF} database.
\newblock In {\em Proceedings of the 2001 IEEE Workshop on the Applications of
  Signal Processing to Audio and Acoustics (Cat. No. 01TH8575)}, pages 99--102.
  IEEE, 2001.

\bibitem{chen2017deep}
Z.~Chen, Y.~Luo, and N.~Mesgarani.
\newblock Deep attractor network for single-microphone speaker separation.
\newblock In {\em IEEE International Conference on Acoustics, Speech and Signal
  Processing}, pages 246--250. IEEE, 2017.

\bibitem{cherry1953some}
E.~C. Cherry.
\newblock Some experiments on the recognition of speech, with one and with two
  ears.
\newblock {\em The Journal of the Acoustical Society of America},
  25(5):975--979, 1953.

\bibitem{dadvar2019robust}
P.~Dadvar and M.~Geravanchizadeh.
\newblock Robust binaural speech separation in adverse conditions based on deep
  neural network with modified spatial features and training target.
\newblock {\em Speech Communication}, 108:41--52, 2019.

\bibitem{domnitz1977lateral}
R.~Domnitz and H.~Colburn.
\newblock Lateral position and interaural discrimination.
\newblock {\em The Journal of the Acoustical Society of America},
  61(6):1586--1598, 1977.

\bibitem{erdogan2015phase}
H.~Erdogan, J.~R. Hershey, S.~Watanabe, and J.~Le~Roux.
\newblock Phase-sensitive and recognition-boosted speech separation using deep
  recurrent neural networks.
\newblock In {\em IEEE International Conference on Acoustics, Speech and Signal
  Processing (ICASSP)}, pages 708--712. IEEE, 2015.

\bibitem{hadad2015theoretical}
E.~Hadad, D.~Marquardt, S.~Doclo, and S.~Gannot.
\newblock Theoretical analysis of binaural transfer function {MVDR} beamformers
  with interference cue preservation constraints.
\newblock {\em IEEE/ACM Transactions on Audio, Speech, and Language
  Processing}, 23(12):2449--2464, 2015.

\bibitem{han2020real}
C.~Han, Y.~Luo, and N.~Mesgarani.
\newblock Real-time binaural speech separation with preserved spatial cues.
\newblock In {\em IEEE International Conference on Acoustics, Speech and Signal
  Processing}, pages 6404--6408. IEEE, 2020.

\bibitem{he2015delving}
K.~He, X.~Zhang, S.~Ren, and J.~Sun.
\newblock Delving deep into rectifiers: Surpassing human-level performance on
  imagenet classification.
\newblock In {\em Proceedings of the IEEE International Conference on Computer
  Vision}, pages 1026--1034, 2015.

\bibitem{hershey2016deep}
J.~R. Hershey, Z.~Chen, J.~Le~Roux, and S.~Watanabe.
\newblock Deep clustering: Discriminative embeddings for segmentation and
  separation.
\newblock In {\em IEEE International Conference on Acoustics, Speech and Signal
  Processing}, pages 31--35. IEEE, 2016.

\bibitem{hershkowitz1969interaural}
R.~Hershkowitz and N.~Durlach.
\newblock Interaural time and amplitude jnds for a 500-{Hz} tone.
\newblock {\em The Journal of the Acoustical Society of America},
  46(6B):1464--1467, 1969.

\bibitem{isik2016single}
Y.~Isik, J.~Le~Roux, Z.~Chen, S.~Watanabe, and J.~R. Hershey.
\newblock Single-channel multi-speaker separation using deep clustering.
\newblock {\em Interspeech}, pages 545--549, 2016.

\bibitem{jensen2016algorithm}
J.~Jensen and C.~H. Taal.
\newblock An algorithm for predicting the intelligibility of speech masked by
  modulated noise maskers.
\newblock {\em IEEE/ACM Transactions on Audio, Speech, and Language
  Processing}, 24(11):2009--2022, 2016.

\bibitem{koizumi2020speech}
Y.~Koizumi, K.~Yaiabe, M.~Delcroix, Y.~Maxuxama, and D.~Takeuchi.
\newblock Speech enhancement using self-adaptation and multi-head
  self-attention.
\newblock In {\em IEEE International Conference on Acoustics, Speech and Signal
  Processing (ICASSP)}, pages 181--185. IEEE, 2020.

\bibitem{kolbaek2017multitalker}
M.~Kolb{\ae}k, D.~Yu, Z.-H. Tan, and J.~Jensen.
\newblock Multitalker speech separation with utterance-level permutation
  invariant training of deep recurrent neural networks.
\newblock {\em IEEE/ACM Transactions on Audio, Speech, and Language
  Processing}, 25(10):1901--1913, 2017.

\bibitem{liu2018iterative}
Q.~Liu, Y.~Xu, P.~J. Jackson, W.~Wang, and P.~Coleman.
\newblock Iterative deep neural networks for speaker-independent binaural blind
  speech separation.
\newblock In {\em IEEE International Conference on Acoustics, Speech and Signal
  Processing}, pages 541--545. IEEE, 2018.

\bibitem{liu2019divide}
Y.~Liu and D.~L. Wang.
\newblock Divide and conquer: A deep {CASA} approach to talker-independent
  monaural speaker separation.
\newblock {\em IEEE/ACM Transactions on Audio, Speech, and Language
  Processing}, 27(12):2092--2102, 2019.

\bibitem{luo2020dual}
Y.~Luo, Z.~Chen, and T.~Yoshioka.
\newblock Dual-path {RNN}: efficient long sequence modeling for time-domain
  single-channel speech separation.
\newblock In {\em IEEE International Conference on Acoustics, Speech and Signal
  Processing}, pages 46--50. IEEE, 2020.

\bibitem{luo2019conv}
Y.~Luo and N.~Mesgarani.
\newblock {Conv-TasNet}: Surpassing ideal time-frequency magnitude masking for
  speech separation.
\newblock {\em IEEE/ACM Transactions on Audio, Speech, and Language
  Processing}, 27(8):1256--1266, 2019.

\bibitem{mandel2009model}
M.~I. Mandel, R.~J. Weiss, and D.~P. Ellis.
\newblock Model-based expectation-maximization source separation and
  localization.
\newblock {\em IEEE Transactions on Audio, Speech, and Language Processing},
  18(2):382--394, 2009.

\bibitem{marquardt2015theoretical}
D.~Marquardt, E.~Hadad, S.~Gannot, and S.~Doclo.
\newblock Theoretical analysis of linearly constrained multi-channel {Wiener}
  filtering algorithms for combined noise reduction and binaural cue
  preservation in binaural hearing aids.
\newblock {\em IEEE/ACM Transactions on Audio, Speech, and Language
  Processing}, 23(12):2384--2397, 2015.

\bibitem{may2011probabilistic}
T.~May, S.~Van De~Par, and A.~Kohlrausch.
\newblock A probabilistic model for robust localization based on a binaural
  auditory front-end.
\newblock {\em IEEE Transactions on Audio, Speech, and Language Processing},
  19(1):1--13, 2011.
\newblock
  \url{http://amtoolbox.sourceforge.net/amt-0.10.0/doc/models/may2011.php}.

\bibitem{nachmani2020voice}
E.~Nachmani, Y.~Adi, and L.~Wolf.
\newblock Voice separation with an unknown number of multiple speakers.
\newblock In {\em International Conference on Machine Learning}, 2020.

\bibitem{reddi2018convergence}
S.~J. Reddi, S.~Kale, and S.~Kumar.
\newblock On the convergence of {Adam} and beyond.
\newblock In {\em International Conference on Learning Representations}, 2018.

\bibitem{reddy2020interspeech}
C.~K. Reddy, V.~Gopal, R.~Cutler, E.~Beyrami, R.~Cheng, H.~Dubey,
  S.~Matusevych, R.~Aichner, A.~Aazami, S.~Braun, et~al.
\newblock The interspeech 2020 deep noise suppression challenge: Datasets,
  subjective testing framework, and challenge results.
\newblock {\em arXiv preprint arXiv:2005.13981}, 2020.

\bibitem{rix2001perceptual}
A.~W. Rix, J.~G. Beerends, M.~P. Hollier, and A.~P. Hekstra.
\newblock Perceptual evaluation of speech quality ({PESQ})-a new method for
  speech quality assessment of telephone networks and codecs.
\newblock In {\em IEEE International Conference on Acoustics, Speech, and
  Signal Processing (ICASSP)}, volume~2, pages 749--752. IEEE, 2001.

\bibitem{stoller2018wave}
D.~Stoller, S.~Ewert, and S.~Dixon.
\newblock {Wave-U-Net}: A multi-scale neural network for end-to-end audio
  source separation.
\newblock In {\em 19th International Society for Music Information Retrieval
  Conference (ISMIR)}, pages 334--340, 2018.

\bibitem{van2007binaural}
T.~Van~den Bogaert, J.~Wouters, S.~Doclo, and M.~Moonen.
\newblock Binaural cue preservation for hearing aids using an interaural
  transfer function multichannel {Wiener} filter.
\newblock In {\em IEEE International Conference on Acoustics, Speech and Signal
  Processing}, volume~4, pages 565--568. IEEE, 2007.

\bibitem{vaswani2017attention}
A.~Vaswani, N.~Shazeer, N.~Parmar, J.~Uszkoreit, L.~Jones, A.~N. Gomez,
  {\L}.~Kaiser, and I.~Polosukhin.
\newblock Attention is all you need.
\newblock In {\em Advances in Neural Information Processing Systems}, pages
  5998--6008, 2017.

\bibitem{venkataramani2018end}
S.~Venkataramani, J.~Casebeer, and P.~Smaragdis.
\newblock End-to-end source separation with adaptive front-ends.
\newblock In {\em 52nd Asilomar Conference on Signals, Systems, and Computers},
  pages 684--688. IEEE, 2018.

\bibitem{wang2018alternative}
Z.-Q. Wang, J.~Le~Roux, and J.~R. Hershey.
\newblock Alternative objective functions for deep clustering.
\newblock In {\em IEEE International Conference on Acoustics, Speech and Signal
  Processing}, pages 686--690. IEEE, 2018.

\bibitem{wang2018end}
Z.-Q. Wang, J.~Le~Roux, D.~L. Wang, and J.~R. Hershey.
\newblock End-to-end speech separation with unfolded iterative phase
  reconstruction.
\newblock {\em Interspeech}, pages 2708--2711, 2018.

\bibitem{wang2019deep}
Z.-Q. Wang, K.~Tan, and D.~L. Wang.
\newblock Deep learning based phase reconstruction for speaker separation: A
  trigonometric perspective.
\newblock In {\em IEEE International Conference on Acoustics, Speech and Signal
  Processing}, pages 71--75. IEEE, 2019.

\bibitem{wightman1992dominant}
F.~L. Wightman and D.~J. Kistler.
\newblock The dominant role of low-frequency interaural time differences in
  sound localization.
\newblock {\em The Journal of the Acoustical Society of America},
  91(3):1648--1661, 1992.

\bibitem{yu2017permutation}
D.~Yu, M.~Kolb{\ae}k, Z.-H. Tan, and J.~Jensen.
\newblock Permutation invariant training of deep models for speaker-independent
  multi-talker speech separation.
\newblock In {\em IEEE International Conference on Acoustics, Speech and Signal
  Processing}, pages 241--245. IEEE, 2017.

\bibitem{zeghidour2020wavesplit}
N.~Zeghidour and D.~Grangier.
\newblock Wavesplit: End-to-end speech separation by speaker clustering.
\newblock {\em arXiv preprint arXiv:2002.08933}, 2020.

\bibitem{zhang2020furcanext}
L.~Zhang, Z.~Shi, J.~Han, A.~Shi, and D.~Ma.
\newblock {FurcaNeXt}: End-to-end monaural speech separation with dynamic gated
  dilated temporal convolutional networks.
\newblock In {\em International Conference on Multimedia Modeling}, pages
  653--665. Springer, 2020.

\bibitem{zhao2020monaural}
Y.~Zhao, D.~L. Wang, B.~Xu, and T.~Zhang.
\newblock Monaural speech dereverberation using temporal convolutional networks
  with self attention.
\newblock {\em IEEE/ACM Transactions on Audio, Speech, and Language
  Processing}, 2020.

\bibitem{zohourian2016binaural}
M.~Zohourian and R.~Martin.
\newblock Binaural speaker localization and separation based on a joint
  {ITD}/{ILD} model and head movement tracking.
\newblock In {\em IEEE International Conference on Acoustics, Speech and Signal
  Processing}, pages 430--434. IEEE, 2016.

\bibitem{zohourian2018gsc}
M.~Zohourian and R.~Martin.
\newblock {GSC}-based binaural speaker separation preserving spatial cues.
\newblock In {\em IEEE International Conference on Acoustics, Speech and Signal
  Processing}, pages 516--520. IEEE, 2018.

\end{thebibliography}

\end{document}